# A spherical Monte-Carlo model of aerosols: Validation and first applications to Mars and Titan

Mathieu Vincendon[1,2] and Y. Langevin[2]

[1]Department of Geological Sciences, Brown University, Providence, RI, USA.

[2]Institut d'Astrophysique Spatiale, CNRS/Université Paris Sud, Bâtiment 121, 91405 Orsay, France (mathieu.vincendon@ias.u-psud.fr).

**Abstract:** The atmospheres of Mars and Titan are loaded with aerosols that impact remote sensing observations of their surface. Here we present the algorithm and the first applications of a radiative transfer model in spherical geometry designed for planetary data analysis. We first describe a fast Monte-Carlo code that takes advantage of symmetries and geometric redundancies. We then apply this model to observations of the surface of Mars and Titan at the terminator as acquired by OMEGA/Mars Express and VIMS/Cassini. These observations are used to probe the vertical distribution of aerosols down to the surface. On Mars, we find the scale height of dust particles to vary between 6 km and 12 km depending on season. Temporal variations in the vertical size distribution of aerosols are also highlighted. On Titan, an aerosols scale height of 80 ± 10 km is inferred, and the total optical depth is found to decrease with wavelength as a power-law with an exponent of −2.0 ± 0.4 from a value of 2.3 ± 0.5 at 1.08 μm. Once the aerosols properties have been constrained, the model is used to retrieve surface reflectance properties at high solar zenith angles and just after sunset.

## 1. Introduction

Mars and Titan's atmospheres are permanently loaded with aerosols. Typical optical depths in the visible and near-IR wavelengths range between 0.2 and 10 as a function of location, season and wavelength (Pollack et al., 1977; Smith et al., 1981; Lemmon et al., 2004; Smith, 2004; Tomasko et al., 2008). Algorithms for separating the surface and aerosol contributions are therefore frequently used when analyzing a Mars or Titan dataset obtained by remote sensing (Clancy and Lee, 1991; Erard et al., 1994; McCord et al., 2006; Rodriguez et al., 2006; Vincendon et al., 2007). The radiative transfer codes widely used in surface–atmosphere separation problems assume a plane-parallel geometry (Stamnes et al., 1988; Evans, 1998). This approach is relevant when the solar zenith angle is small enough to neglect the curvature of the planet (typically less than 80° for Mars and 65° for Titan, which has a larger relative scale height). Beyond that limit, the use of spherical codes is required (Collins et al., 1972; Adams and Kattawar, 1978; Herman et al., 1994). The high solar zenith angles encountered around the terminator increase the effective optical depth of aerosols, which prevents the use of simplifying single scattering approximations such that followed by Santer et al. (1986). So far, only Monte-Carlo methods are able to take into account the planetary curvature without simplifying assumptions on multiple scattering (Blättner et al., 1974; Oikarinen et al., 1999; Tran and Rannou, 2004; Pitman et al., 2008). However, spherical Monte-Carlo algorithms are time-consuming and have been mainly used for limited planetary data processing (Whitney et al., 1999; Lebonnois and Toublanc, 1999; Vasilyev et al., 2009) or for validation purpose (Tran and Rannou, 2004; Tomasko et al., 2008).

In this paper we describe and implement a fast 3D Monte-Carlo algorithm of radiative transfer through particles in a spherical, laterally homogeneous atmosphere and include the surface–atmosphere interface. Our approach takes advantage of the symmetry of homogeneous spheres by reducing the problem to a 1D atmosphere and surface (but with 3D photometric/radiative transfer effects) in order to reduce the computational time. This modeling approach can be compared to observations when the "independent pixel approximation" is relevant, i.e. when all photons collected by an instrument looking at a given location are the result of radiative transfer in atmospheric and surface properties representative of this location. This requires that aerosols and surface properties do not significantly vary over lateral distances from one to several atmospheric scale heights (i.e., typically around 10 km for Mars and 60 km for Titan). The first section of this paper describes the equations of the Monte-Carlo algorithm. The model is then applied to observations around the terminator obtained by OMEGA/Mars Express and VIMS/Cassini. The observations are used to study the vertical distribution of aerosols down to the surface, which is not possible with the more widely used limb observations for which the aerosol layer appears optically thick close to the surface. Our spherical algorithm is then used to extend surface reflectance retrievals to seasons and latitudes where the surface is only illuminated by high solar zenith angles.

## 2. A 3D spherical Monte-Carlo algorithm

### 2.1. Assumptions and symmetries

We consider a spherical planetary body with the surface defined by its local albedo $A$ (the fraction of photons reaching the surface that are scattered toward space), and its angular scattering law. Three parameters describe the aerosol layer: the single scattering albedo and the single scattering phase function of the particles, and the optical depth of the whole aerosols layer. These properties do vary with altitude which is represented by the atmosphere being composed of concentric layers ("a spherical shell atmosphere"). Positions are referenced using an orthonormal basis { $\vec{x}$, $\vec{y}$, $\vec{z}$ }. The { $\vec{x}$ } axis links the Sun and the center of the planetary body and corresponds to the direction of incoming solar photons. The { $\vec{y}$, $\vec{z}$ } plane corresponds to the terminator plane. The system is symmetrical about the $\{\vec{x}\}$ axis. Under the assumption of lateral homogeneity, test photons can be sent in the $\{\vec{x}, \vec{z}\}$ plane only, surface elements being defined as rings centered on the symmetry axis $\{\vec{x}\}$. Incoming solar photons will then travel in the 3D environment after being scattered outside their initial $\{\vec{x}, \vec{z}\}$ plane. All the notations mentioned in this paragraph are described in Fig. 1.

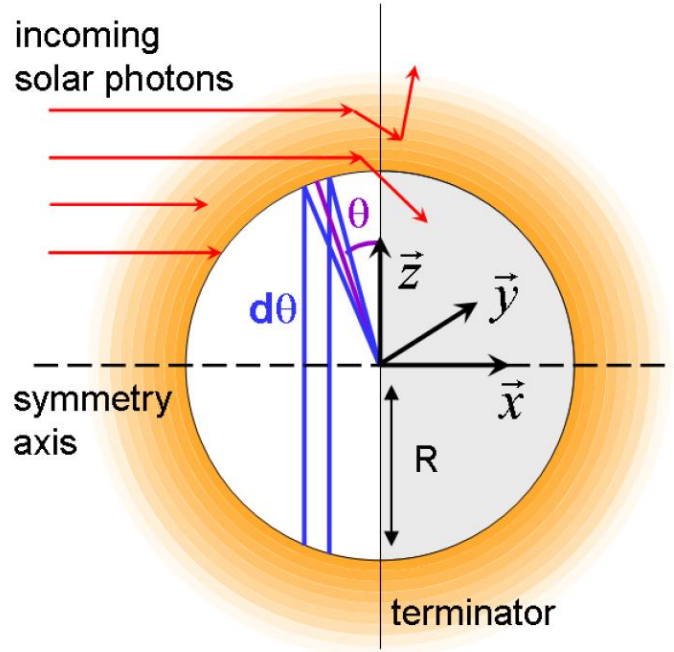

**Figure 1.** *3D spherical geometry with notations. R is the radius of the planet. Aerosols are represented by concentric layers with a decreasing optical depth as a function of the altitude.*

### 2.2. Probability linked with single paths

We first create a 2D pre-calculated array that contains the optical depths associated with a set of paths from which it will be possible to infer the optical depth linked with any path in the 3D spherical atmosphere. The geometry of this array is described in Fig. 2. We will detail in Section 2.3, how to link any path in the 3D environment with paths from this array.

The array contains two types of paths: type "A" that do not intersect with the planet, and type "B" that do (see Fig. 2). Each "line" (i.e., along the $\{\vec{x}\}$ axis) of the array is filled by integrating the optical depth along the corresponding path, starting from the middle of the atmosphere for type "A" paths and starting from the surface for type "B" paths. For example (type "A" path), the value contained in the array at coordinates $\{z_0, x_0\}$ is given by:

$$\tau_{array}(z_0, x_0) = \int_0^{x_0} \tau_{normalized}(x) dx$$

with $x = -\sqrt{(z'^2 - z_0^2)}$, $z'$ is the altitude and $\tau_{normalized}$ is the optical depth per unity of length.

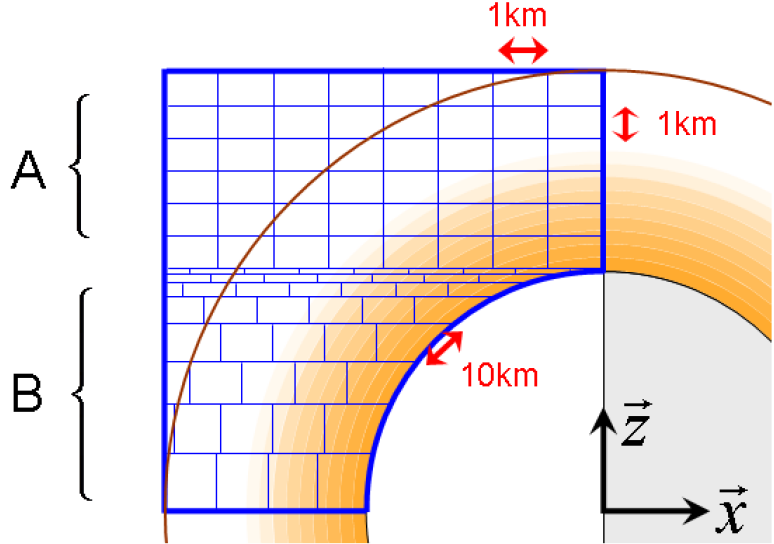

**Figure 2.** *Geometry of the pre-calculated array that contains all the possible single paths (the single paths are along the x axis). All actual photon routes in the 3D environment are combinations and interpolations of the paths contained in this array. Two types of single paths, noted "A" and "B", constitute the array. Paths "A" are in the aerosols layer only (above the limb) while paths "B" intersect the surface. The typical spatial sampling around which no perceptible impacts on accuracy are observed (see Fig. 7d) is one line every kilometer of altitude for "A" paths and one line every 10 km along the surface for "B" paths. Each path is sampled every kilometer (x axis). The summit of the aerosols layer is delimited by the brown line: it corresponds to the altitude above which the optical depth and therefore the probability of interaction with aerosols are negligible (typically 100 km for Mars, 500 km for Titan). The values contained in the array correspond to a single direction (from right to left); the optical depths of the opposite direction can be simply obtained by subtracting the total optical depth along the line of interest.*

### 2.3. Event loop

Once this array is filled, we can send photons in the system and follow their routes in the 3D environment by using random numbers to sample the probability of the different possible events (Hammersley and Handscomb, 1964; Brinkworth and Harris, 1969): probability of interaction (given by the optical depths of the array), probability of scattering versus absorption (given by the albedos), and directional probability when a scattering event occurs (given by the phase function of aerosols and the angular scattering law of the surface). The photons are positioned in the { $\vec{x}$, $\vec{y}$, $\vec{z}$ } basis with coordinates { $X$, $Y$, $Z$. A second orthonormal basis { $\vec{u}_x$, $\vec{u}_y$, $\vec{u}_z$ } is associated with each photon, where { $\vec{u}_z$ } is colinear to the direction of the photon. This basis will therefore change as the direction of the photon change. We initially send a flux $F$ of solar photons ($F$ is a number of photons per square kilometer). The photons start at coordinates $\{X = -(R+H_{top}),\ Y = 0,\ Z \in [0; R+H_{top}]\}$, where $H_{top}$ is the "summit" (or top) of the atmosphere (brown line in Fig. 2). As the system is symmetrical about the $x$ axis, photons can be sent from a single line parallel to the $z$ axis. The number of photons sent at an altitude $z$ corresponds to the flux that goes through the ring surface element $dS = 2\pi z dz$ ($N_{photon}(z) = FdS$). After the first part of the simulation, a photon can either be absorbed by aerosols, escape the planet, or reach the surface. The surface of the planetary body is divided into ring surface elements centered on the axis of symmetry (the $x$ axis): $dS_{surface} = 2\pi R^2 \cos(\theta) d\theta$ (Fig. 1).

In order to determine whether a given photon interact with aerosols during a given path, and where the potential interaction occurred, we need to know the optical depth that is in front of the photons, and the distribution of this optical depth with distance. This is tabulated in the pre-calculated array (Fig. 2). To know which path of the array need to be considered, we need to determine the impact parameter "$b$", the distance "$d$" between the initial position of the photon and the origin of the array (middle of the atmosphere for "A" paths, surface for "B" paths), and the direction of the photon in the array (from right to left, or the opposite). Corresponding notations are shown in Fig. 3 and Fig. 4.

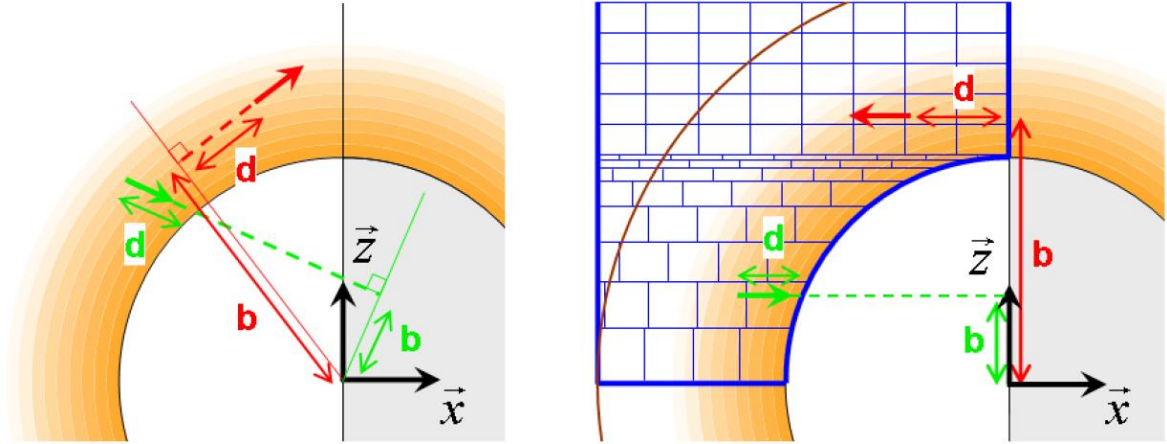

**Figure 3.** *Relation between real paths in the 3D environment (left) and corresponding paths in the pre-calculated array (right). Two examples of photons are shown: one going toward the planetary body (green arrow, "B" path) and one going through the aerosols layer above the surface (red arrow, "A" path). "b" is the impact parameter and "d" is the distance between the initial position of the photon and the limit of the array (the plane perpendicular to the direction of the photon that contain the center of the planetary body for "A" paths, and the distance to the planetary body for "B" paths). The direction of propagation of these two photons in the array is opposite (green: from left to right; red: from right to left).*

We first calculate the algebraic distance $D$ between the position $M$ of the photon and the plane $P$ perpendicular to $\{\vec{u}_z\}$ that contain the center $C$ of the planetary body (see Fig. 4 for notations). We note that $\{\vec{u}_z(0), \vec{u}_z(1), \vec{u}_z(2)\}$ is the coordinate of $\{\vec{u}_z\}$ in $\{\vec{x}, \vec{y}, \vec{z}\}$. From $\vec{CM} \cdot \vec{u}_z = \|\vec{CM}\|\cos(\theta)$ and $D = \|\vec{CM}\|\cos(\theta)$ we obtain:

$$D = X\vec{u}_z(0) + Y\vec{u}_z(1) + Z\vec{u}_z(2)$$

The direction of propagation of the photon in the array is given by the sign of $D$: if $D > 0$, the photon keeps away from the impact plane $P$ (it goes from right to left). The impact parameter $b$ is given by $b = \|\vec{CB}\|$ with $B$ defined as the closest point to $M$ in plane $P$:

$$\{X_B = X - D\vec{u}_z(0),\ Y_B = Y - D\vec{u}_z(1),\ Z_B = Z - D\vec{u}_z(2)\}$$

If $b > R$, the trajectory of the photon does not intersect the planetary body (route "A") and $d = \|D\|$. If $b \leq R$, the trajectory of the photon crosses the planetary body and $d = \|D\| - \sqrt{R^2 - b^2}$.

The three parameters "$b$", "$d$", and the direction, makes it possible to find (from interpolation) the distribution of optical depth as a function of the distance that is in front of the photon. We then draw a random number $n$ between 0 and 1, which is converted to an optical depth $tn$ using the relationship $tn = -\ln(n)$. By comparing $tn$ to the distribution of optical depth that is in front of the photon, it is possible to determine whether or not there is an interaction, and if so where along the path the interaction occurs:

(1) In the case of a scattering event, it is necessary to determine the new propagation direction. This new propagation direction is defined by two angles: the scattering angle $\theta$ and the azimuth $\varphi$. The distribution of azimuth is equi-probable (between 0 and $2\pi$), while the scattering angle $\theta$ must be determined through sampling the single scattering phase function of aerosols $P(\theta)$. Such a random number selection can be obtained by drawing a number between 0 and 1. This number is then converted using the relation between $\theta$ and $\int_0^\theta P(\theta')d\theta'$, which ranges from 0 to 1. The angles $\theta$ and $\varphi$ makes it possible to determine the new basis $\{\vec{u}_{x2}, \vec{u}_{y2}, \vec{u}_{z2}\}$ associated with the photon with respect to the previous one $\{\vec{u}_x, \vec{u}_y, \vec{u}_z\}$, as follows (see also Fig. 5).

$$\vec{u}_{z2} = \cos(\theta)\vec{u}_z + \sin(\theta)(\cos(\varphi)\vec{u}_x + \sin(\varphi)\vec{u}_y)$$
$$\vec{u}_{x2} = -\sin(\theta)\vec{u}_z + \cos(\theta)(\cos(\varphi)\vec{u}_x + \sin(\varphi)\vec{u}_y)$$
$$\vec{u}_{y2} = \cos(\varphi)\vec{u}_y - \sin(\varphi)\vec{u}_x$$

(2) If a photon is on a type "A" path (the surface of the planetary body is not on its trajectory) toward the right limit of the array, and if it reaches this limit, then we remove to the optical depth *tn* the optical depth already covered, we reverse the direction of propagation in the array, and we repeat the process starting from the beginning of the array.

(3) If a photon reaches the surface, we calculate the coordinates $\{X_0, Y_0, Z_0\}$ of the intersection point $M_0$ between the trajectory and the planet. It is given by the equation $\vec{CM}_0 = \left(R/\|\vec{CM}\|\right) \times \vec{CM}$. Next we determine the angle $\theta$ between the terminator plane and the intersection surface element (see Fig. 1), $\theta = \pi/2 - \arccos(X_0/R)$ (positive $\theta$ angles correspond to surface elements that are on the night side). This is used to calculate the flux receive by the surface.

(4) If the photon leaves the aerosols layer toward space on a "B" path (i.e., path intersecting the surface), it contributes to the radiance as seen by an instrument observing the surface. At this point, we need to determine from which surface element the photon seems to originate. We calculate as before the coordinates $\{X_0, Y_0, Z_0\}$ of the intersection point between the trajectory and the planetary body. The associated emergence angle is derived from these coordinates and the values of $\{\vec{u}_x, \vec{u}_y, \vec{u}_z\}$.

(5) If the photon escapes the aerosols layer on an "A" trajectory, we need to determine the altitude "h", the solar zenith angle "i", and the azimuth angle (see Fig. 6). This photon would be observed by an instrument performing a limb observation.

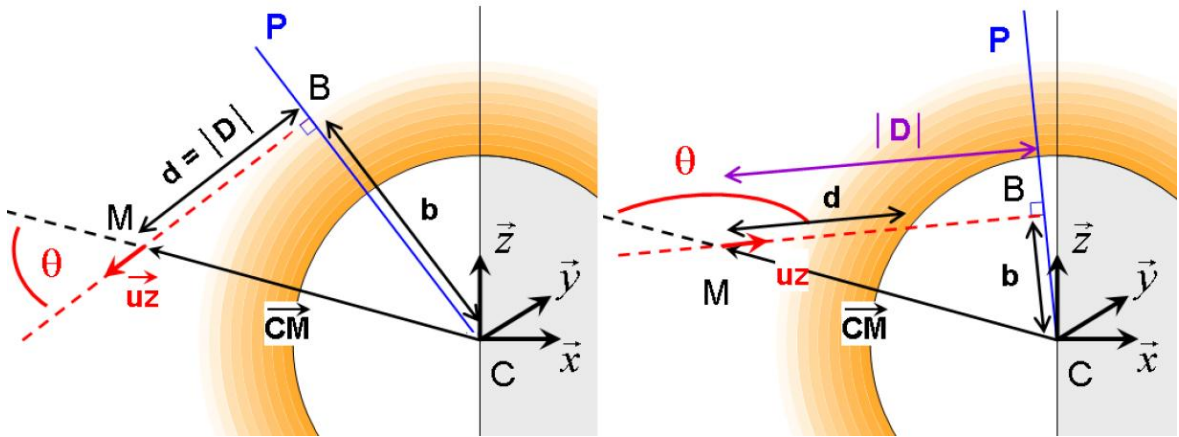

**Figure 4.** *Notations used for the calculation of the parameters "d" and "b" for (a) photons which trajectory do not intersect with the planetary body (type "A") and (b) photons which trajectory intersect the planetary body (type "B").*

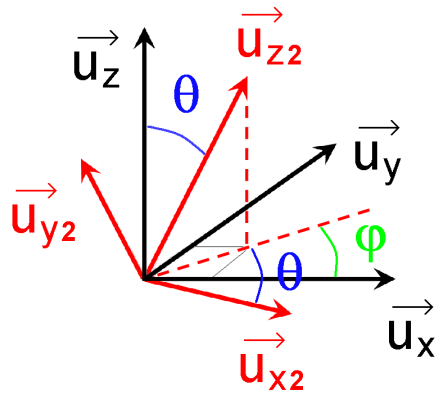

**Figure 5.** *Relation between the two bases linked with a photon before and after a scattering event.*

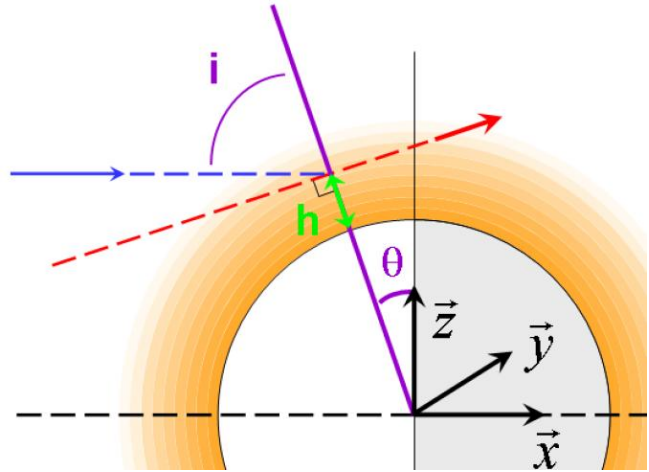

**Figure 6.** *Photon going out of the aerosols layer above the limb of the planetary body (red arrow). For this photon we need to calculate the solar zenith angle "i" compared to the initial direction prior to the last interaction (blue arrow), the altitude "h" (similar calculation as the impact parameter "b" in Fig; 5), and the azimuth.*

### 2.4. Building look-up tables of observed reflectance

As described above, our approach provides the distribution of photons that escape the planetary body after scattering by aerosols and before surface interaction. From this the associated reflectance "$I/F_{aerosols}$" for a given geometry can be calculated using the approach of Vincendon et al. (2007). The model also provides the fraction of the flux that reaches the surface, $I/F_{surface}$. The number of photons scattered by the surface of albedo $A$ is then given by $A \times I/F_{surface}$. This irradiance is modified by the aerosols layer as it again passes through the atmosphere. This portion of a photon trajectory can be modeled using a plane-parallel code (e.g. Vincendon et al., 2007) if the observations are limited to moderate emission angles, or if not using the spherical code (angles typically greater than 60°). This second simulation provides a transmission factor *T*, as well as the fraction of photons emitted by the surface that goes back to the surface $I/F_{back}$. The normalized radiance (a.k.a., radiance factor or *I/F*) for a given set of parameters is then calculated by combining the results of these two simulations with the following equation:

$$I/F = I/F_{aerosols} + I/F_{surface} \times A \times T \times \left(\frac{1}{1 - A \times I/F_{back}}\right)$$

where the infinite sum of a geometric series (term in bracket) arises from the successive back and forth movements of photons emitted by the surface and scattered back toward the surface by aerosols.

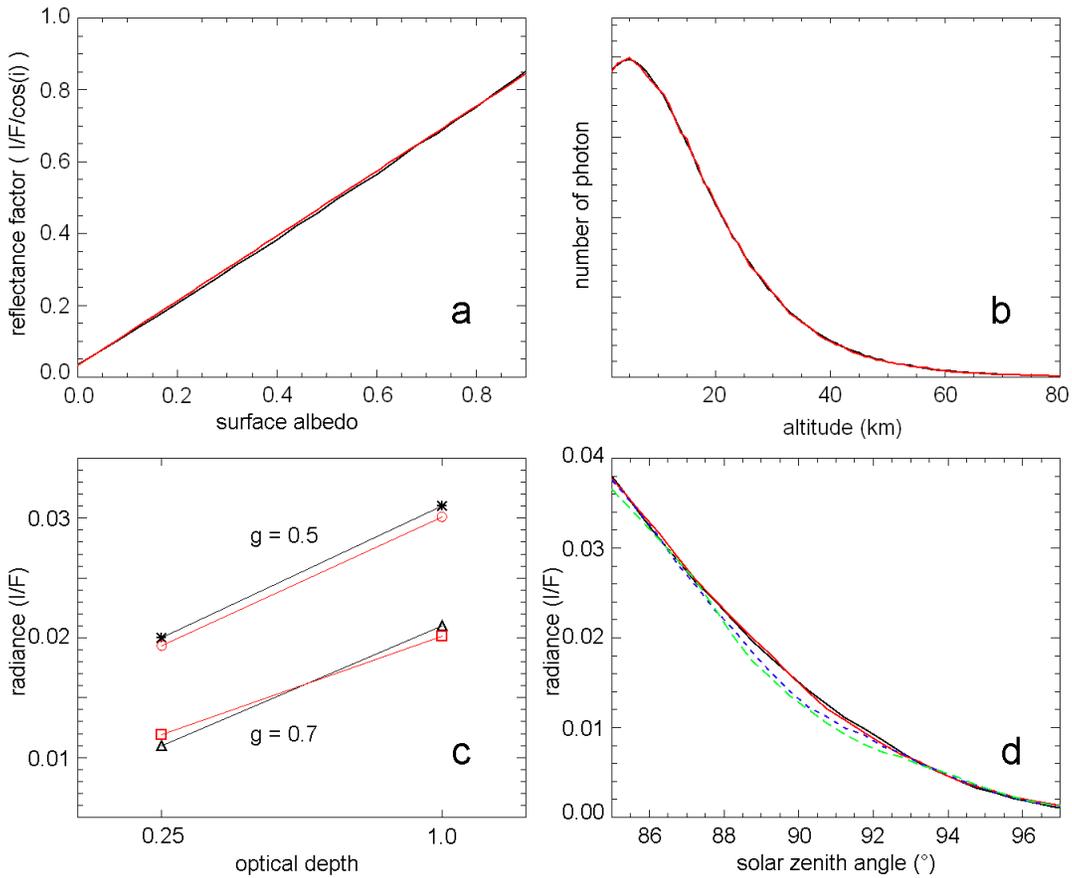

**Figure 7.** *Model validations. Panels a and b: comparison between the 3D spherical code (red) and the plane-parallel code of Vincendon et al. (2007) (black), in the common validity domain of the two models. (a) Variations of the apparent reflectance in a nadir viewing geometry as a function of the surface albedo for an optical depth of 0.2 and a solar zenith angle of 66°. (b) Number of photons that reach the surface as a function of the altitude of the first interaction with aerosols for a solar zenith angle of 75° and an optical depth of 0.5. A total of $4 \times 10^6$ photons are used for the plane parallel model, $2 \times 10^7$ for the*

*spherical model. The two models give similar results. Panel c: comparison between the 3D spherical code (red, circles or squares) and the results computed with another spherical model as published by Kattawar and Adams (1978) (black, stars or triangles). A solar zenith angle of 84.25° is used. The predicted radiances of the aerosols layer are compared for two values of the optical depth and of the asymmetry parameter g of the aerosols phase function. Both spherical models predict observed radiances which differ by only a few %. (d) Sensitivity of the model results to the spatial sampling of the pre-calculated array (Fig. 2). The radiance as a function of the solar zenith angle is shown for four samplings: case 1 (black line), sampling of Fig. 2 (1 km/10 km/1 km); case 2 (red line), sampling two times lower (2 km/20 km/2 km); case 3 (blue dotted line) and case 4 (green dashed line), sampling four times and eight times lower respectively. Similar model results are obtained for a high sampling (cases 1 and 2). Model results are affected around the terminator if the sampling is too low (cases 3 and 4). Martian properties are used for panels a,b and d (see Table 1). Earth atmosphere properties as described in Kattawar and Adams (1977) are used in panel c. An emission angle of 0 (nadir pointing) is considered for all panels.*

This approach makes it possible to limit the number of simulations. Rather than computing all the radiative transfer for each surface albedo, the simulation is divided in two parts: from the Sun toward the surface and from the surface toward the spacecraft. The results are then combined using the above equation to provide the exact solution for all possible surface albedo value, *A*.

### 2.5. Validation

Our spherical Monte-Carlo code has been validated using the common validity domain of spherical and plane-parallel codes. In Fig. 7a and b, two examples of results of the spherical code are compared to that of the plane parallel Monte-Carlo code of Vincendon et al. (2007). Both models give similar results at moderate incidence angles. In addition, we have performed some comparisons between our model and the spherical model of Kattawar and Adams (1978) for a high solar zenith angle. Both spherical models predict observed radiances which differ by a few % only (Fig. 7c).

In Fig. 7d, we study the sensitivity of the model results to the sampling of the pre-calculated array. We compare model results obtained with the spatial sampling described in Fig. 2 (case 1) with samplings decreased by a factor of 2, 4 and 8 (cases 2, 3 and 4 respectively). Results begin to be affected by the sampling when it is reduce by a factor of 4 or more. Given the limited time gain between cases 1 and 2 (about 5–10% depending on optical depth), we have selected the conservative sampling of case 1.

### 3. First applications to observations of Mars and Titan at the terminator

### 3.1. Observations

In this section, we analyze at set of observations of Mars obtained by the *Observatoire pour la Minéralogie*, *l'Eau*, *les Glaces et l'Activité* (OMEGA) onboard Mars Express and observations of Titan performed by the *Visual and Infrared Mapping Spectrometer* (VIMS) onboard Cassini. OMEGA and VIMS are imaging spectrometers observing in the 0.3–5.1 µm wavelength range. Their spatial resolution ranges from a few hundreds of meters/pixel to a few kilometers/pixel. The selected observations were obtained close to the terminator: solar zenith angles are greater than 70° and include in some case a portion of the night side. These observations have been obtained in either a nadir pointing geometry or with moderate emission angles (<30°). We focus on the near-IR wavelengths that do not contain strong gas absorption bands. This corresponds to wavelengths of 1.08 µm, 1.28 µm, 1.60 µm and 2.03 µm for VIMS and of the (1 µm–2.6 µm) range for OMEGA, where for the latter the gas absorptions can be corrected using the method of Langevin et al. (2007).

### 3.2. Modeling assumptions

We choose to approximate the surface scattering law by a Lambert law. This simplification is relevant if specular and anti-solar phase angles are not considered (for Mars, see e.g. Johnson et al., 2006), particularly given the large fraction of diffuse light present in our observations (Vincendon et al., 2009). The single scattering parameters for the aerosols in the near-IR (albedo and phase function) have been constrained by previous studies (Ockert-Bell et al., 1997; Tomasko et al., 1999; Tomasko et al., 2008; Vincendon et al., 2007; Määttänen et al., 2009; Wolff et al., 2009). On Titan, those properties have been reported to vary slightly with altitude (Tomasko et al., 2008). On Mars, while it is expected that they also are a function of altitude, many aerosol retrievals have generally assumed them to be constant especially when one is observing the entire aerosol layer (e.g., Wolff et al., 2009). For simplicity, we adopt the same assumption. The employed values are summarized in Table 1. The vertical structure of the optical depth will be represented by a constant scale height in an exponential atmosphere $\tau_V(z) = \tau \exp(-z/H)$. On Titan, this model provides a first approximation of the observed vertical variations at wavelengths greater than 1 µm (Tomasko et al., 2008, Fig. 50). On Mars, this hypothesis relies on two main underlining assumptions: that dust is "well-mixed" with gas and that the temperature of the atmosphere is constant with altitude. Dust is usually observed to be well-mixed in the bottom scale height or two, where one typically find most of the optical depth (Lemmon et al., 2004; Zasova et al., 2005), and temperature variations over that altitude range are smaller than 20% (Smith, 2008). Each observation is thus characterized at each wavelength by three free parameters: the surface albedo ($A$), the aerosols optical depth ($\tau$), and the scale height of aerosols ($H$). For each planet we built a look-up table of apparent reflectance as a function of these three parameters. Surface albedo varies from 0.04 to 0.92 with a sampling of 0.04, aerosol optical depths are 0.1, 0.3, 0.5, 0.7, 1.0, 1.3, 1.6, 2.0 and 2.5, and scale height varies every kilometer between 6 and 12 km for Mars and every 5 km between 45 and 90 km for Titan. The radius is set to 3380 km for Mars and 2575 km for Titan.

In the next sections, we analyze observations composed of several radiance measurements at different solar zenith angles of the same {surface–atmosphere} system. Using interpolated version of the look-up tables with higher samplings, we look for the set of parameters {$A$, $\tau$, $H$} that provides the best model fit with a simple least square approach (the same weight is attributed to all radiance measurements). We will either constrain the three parameters at the same time or fix two parameters using other observational constraints and let vary the last remaining parameter. Uncertainties on the retrieved parameters are estimated using the range over which the root mean square increases by two compared to its minimum "best fit" value (for multi-parameters fit, we estimate the uncertainty parameter by parameter while the others remain at their "best fit" value)".

**Table 1:** Selected optical parameters for aerosols in the near-IR (1 µm – 2.5 µm)

| Single scattering albedo | Phase function |
|---|---|
| *Mars* | |
| 0.974 (Vincendon et al., 2007) | Henyey–Greenstein, asymmetry of 0.63 (Ockert-Bell et al., 1997) |
| *Titan* | |
| 0.984 (1.1 µm), 0.974 (1.3 µm), 0.954 (1.6 µm), 0.93 (2.0 µm) (Tomasko et al., 2008) | Phase functions from Tomasko et al. (2008), Table 1a |

### 3.3. Mars

#### 3.3.1. Retrievals of the aerosol scale height

Two observations of Mars obtained around the terminator are illustrated in Fig. 8. The observation in Fig. 8a was obtained in the southern hemisphere at $L_S$ 301°, during the planet-encircling dust event of 2007. The optical depth at that time is expected to be high over much of the planet. From MER measurements (Mark Lemmon, personal communication) it is expected to be equal to 1.6 once scaled for elevation of the OMEGA observation. Attempting to retrieve the dust scale height leads to a good model fit to the observation. The surface albedo is determined independently from other observations of the same region but obtained with a lower aerosols optical depth and a lower solar zenith angle: $A = 0.19$. We find a scale height of 11.5 km ± 1 km, which is in agreement with the values previously derived for similar conditions (Lemmon et al., 2004; Zasova et al., 2005). The dust is expected to be well-mixed to altitudes high in the atmosphere as this observation is obtained during a dust storm and during summer, when the atmosphere is warm. The observation in Fig. 8b is obtained at polar latitudes (80°S) in late summer ($L_S$ 340°) of 2004. Solving for the three unknowns simultaneously leads to reasonable parameters: a surface albedo of 0.13, and aerosols optical depth of 1.59 (a value consistent with the measurements of Smith (2004)), and a scale height of 6 ± 2 km. The difference between the two observations is consistent with the expected atmospheric temperature change (Smith, 2004). We must however notice that other solutions not explored in details here would also provide satisfactory results. Other profiles for which the dust is confined close to the surface would also works for the observation of Fig. 8b. For example, dust could be mixed with a scale height around 10 km and then dramatically reduced above 10–15 km because of water ice scavenging processes.

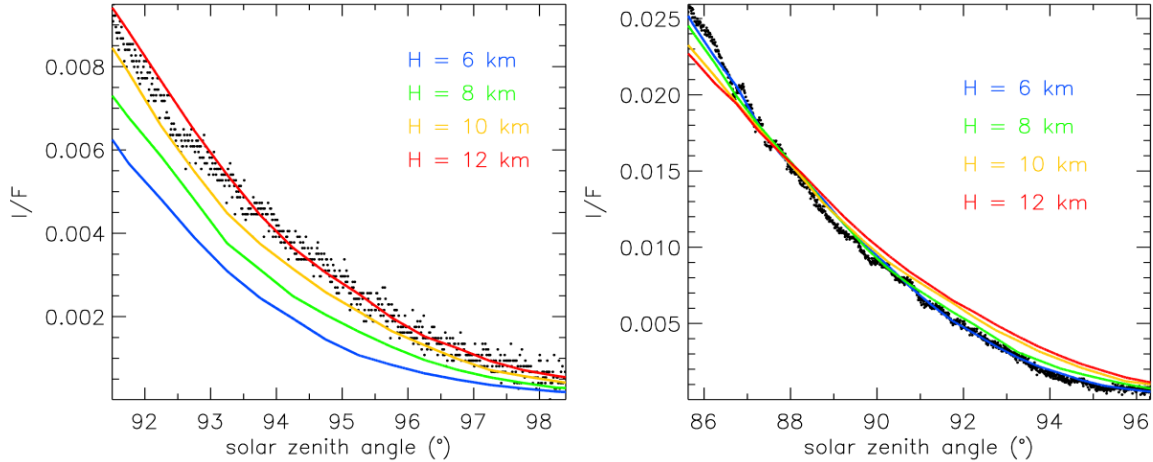

**Figure 8.** *Black: reflectance at 1.1 μm measured by OMEGA. Model results are indicated in colors for different dust vertical scale heights H. (a) Observation in the night side at $L_S$ 301° in August 2007, covering the region between 56°S and 64°S at 222°E. Model results are indicated for a surface albedo of 0.19 and an optical depth of 1.6. (b) Observation obtained at 80°S, 280°E in 2004 ($L_S$ 338°). The solar zenith angle range is wider and includes both day and night. All parameters (optical depth, surface albedo, dust scale height) are free. The retrieved surface albedo and optical depth are 0.13 and 1.59 respectively. A smaller dust scale height is required to explain this observation obtained close to the south pole and autumn equinox.*

3.3.2. Particle size variations

Several studies have pointed out the existence of a decrease in the particle size of the aerosols with altitude (Chassefiere et al., 1992; Korablev et al., 1993; Rannou et al., 2006; Montmessin et al., 2006). Spatial and seasonal variations in the amplitude of that gradient are also expected from dynamical modeling results (e.g. Kahre et al., 2008). In the near-IR, a change of particle size primarily modifies the spectral slope of the single scattering cross section (Drossart et al., 1991; Clancy et al., 2003; Vincendon et al., 2009). The smaller the particles are, the steeper the spectral slope is. The average altitude of the first scatter with aerosols increases with incidence angle (Fig. 9). Therefore, we look for such variations in the spectral slopes in the terminator observations as an indication of particle size variations with altitude. Because the aerosols cross section is directly proportional to the optical depth, a change in particle size with altitude can therefore be modeled by a wavelength dependent scale height. A constant scale height corresponds to a constant particle size with altitude, while a decreasing scale height is equivalent to an increasing optical depth slope with altitude, i.e. a decreasing particle size. Two observations showing strong differences in terms of spectral slopes variations are highlighted in Fig. 10 and Fig. 11. In Fig. 10, a constant scale height provides a good fit to the observation, suggesting that no major variations of the particle size occur within the first tens of kilometers of the atmosphere. The observation in Fig. 11 requires a decrease of the scale height by 5% between 1 μm and 2.5 μm. This corresponds to an increase of the ratio of optical depth $\tau$ (1 μm)/$\tau$ (2.5 μm) by 10% every 10 km. This corresponds to a vertical gradient of particle size of the order of 0.01–0.02 μm km$^{-1}$ according to T-matrix modeling (Mishchenko, 1998; Vincendon et al., 2009). This observation was obtained at the spring equinox above the south polar cap when a thin layer of water ice clouds are observed in the atmosphere (Langevin et al., 2007). The presence of these clouds may contribute to the apparent decrease of particle size with altitude.

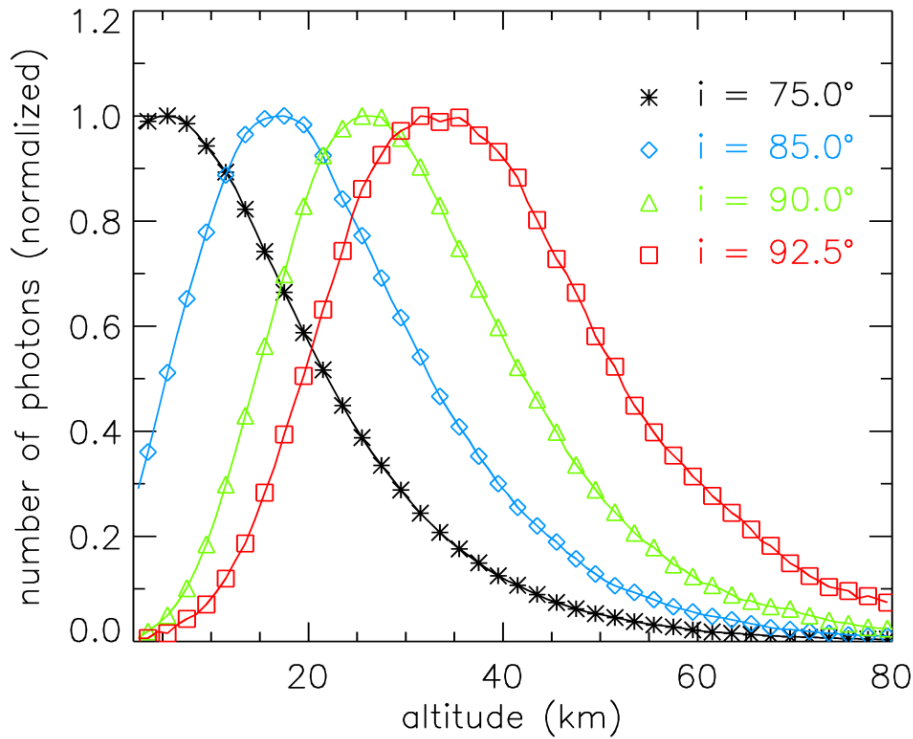

**Figure 9.** *Altitudes at which photons that reach the surface have been scattered by aerosols, for four solar zenith angles i. Aerosols properties corresponding to martian dust are used (see Table 1; the dust scale height is 11.5 km). The total optical depth is 0.5. The altitude of interaction of photons increases with the solar zenith angle. At ±3° of the terminator, photons that are collected by the instrument have been first scattered high in the atmosphere (~20–40 km).*

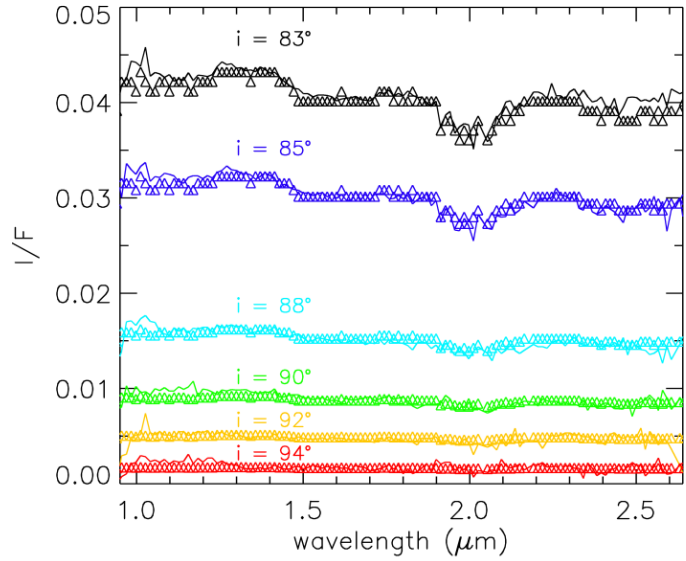

**Figure 10.** *Observed near-IR spectra extracted from the observation of Fig. 8b (thin lines) for six solar zenith angles (from top to bottom, 83°, 85°, 88°, 90°, 92°, 94°). These observations can be reproduced with the model (symbols) with an optical depth of 0.19 at 1 μm that linearly decreases to ~ 0.1 at 2.5 μm, a scale height of 6 km for all wavelengths and a surface albedo spectrum extracted from a previous observation at low aerosols contribution.*

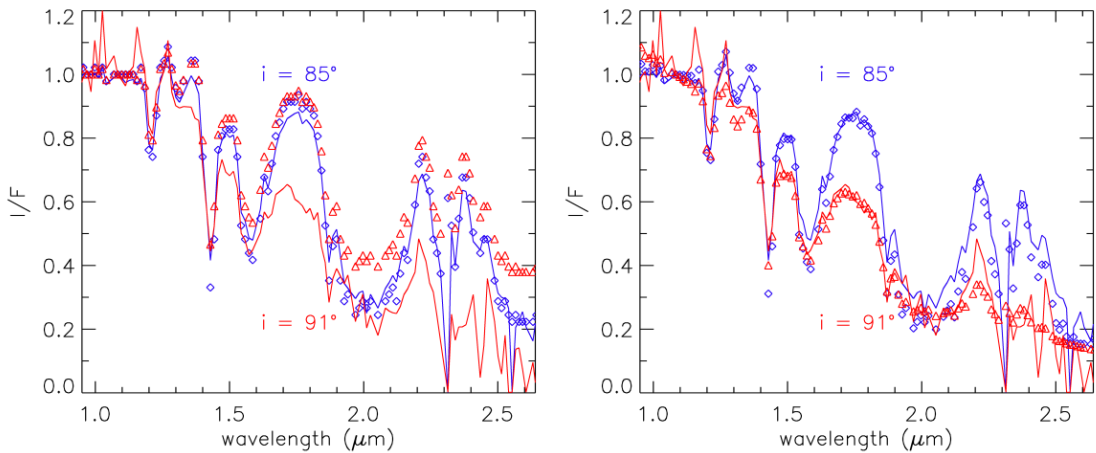

**Figure 11.** *A similar $CO_2$ ice surface is observed at i = 85° (blue diamonds) and at i = 91° (red triangles). The observation has been obtained in the south polar regions of Mars at $L_S$ 175°. Observations are in thin solid line, and model results in symbols. Two different hypotheses regarding the scale height are implemented: (a) constant scale height of 6 km; (b) the scale height linearly decreases from 6 km at 1 μm to 5.7 km at 2.5 μm. Observations cannot be simultaneously reproduced with a constant scale height (Fig. 11a), while a decreasing scale height with wavelength provides a better match (Fig. 11b).*

### 3.3.3. Surface retrievals

In Section 3.3.1, we were able to simultaneously retrieve the surface albedo and the aerosol properties because we selected observations for which the surface albedo is approximately constant for the observed range of solar zenith angles. To retrieve a surface albedo map of an inhomogeneous region observed around the terminator, we need to first constrain the aerosols properties of such a region, and then to apply the model over a broader region by assuming no major change in the aerosol properties. An example of this approach is shown in Fig. 12. The selected data was obtained a few days before the southern fall equinox, during which time the Sun is a few degrees above or below the horizon at the highest latitudes. We first retrieve the aerosols properties by focusing on the larger solar zenith angles of the observations (similarly to Fig. 8a), as the aerosol contribution dominates in that range: an optical depth of 0.1 and a scale height of 6 km are derived. These parameters are then used to retrieve the albedo map of the surface for all elements in the observation. As shown by Fig. 12, the surface albedos as corrected for the aerosol contribution are quite consistent with those at lower incidence angles (obtained earlier in the season, but during other martian years). This suggests that no major interannual or seasonal changes appear to have occurred in that region.

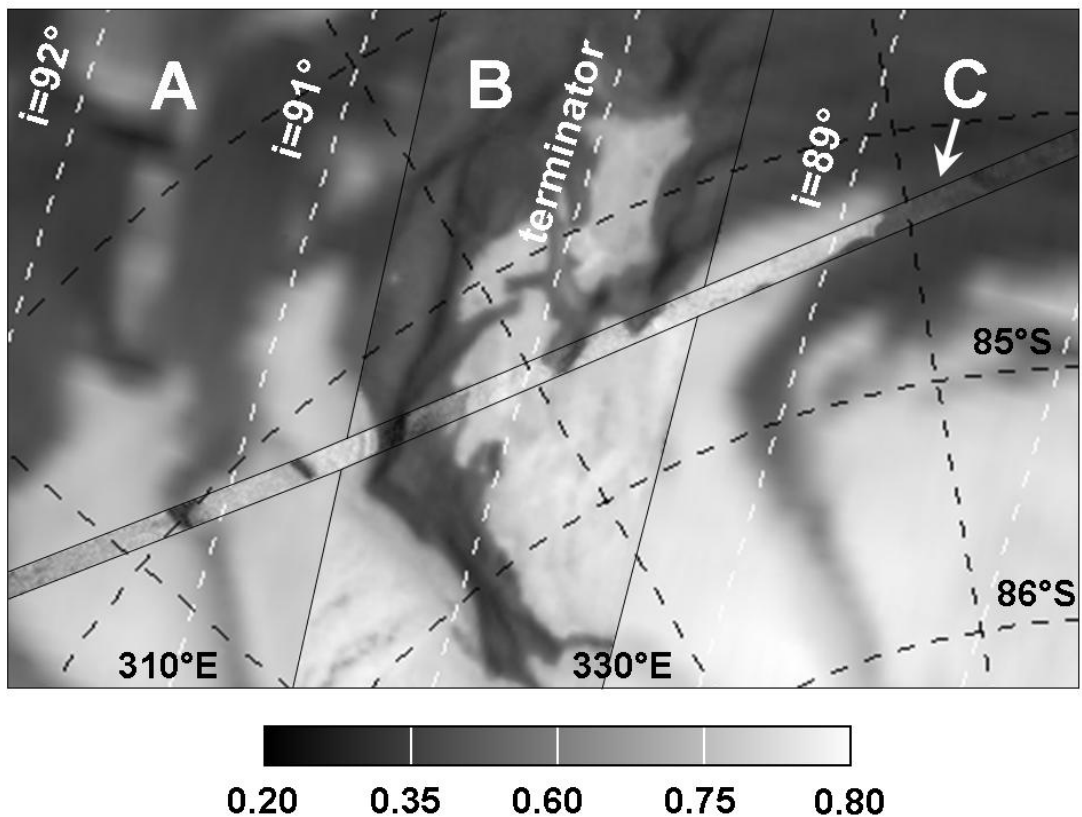

**Figure 12.** *Surface albedos for three observations of a region of the south perennial cap of Mars between 84°S and 85°S obtained in late summer. The background image (A) has been obtained at a resolution of 4 km/pixel at $L_S$ 335° in 2009. An observation (B) at a higher*

*resolution (1 km/pixel) that covers the central part of the image has been obtained at $L_S$ 335° one martian year earlier (2007). These two images have been obtained with incidence angles around 80°; the surface albedo has been retrieved using the plane-parallel aerosol correction model of Vincendon et al. (2008). The narrow high-resolution track (C) has been obtained at LS 352° in 2005 across the terminator with a resolution of ~300 m/pixel. The surface albedos for this observation at very high incidence angles (white dashed lines) have been determined with the spherical Monte-Carlo model presented in this contribution assuming a vertical optical thickness of 0.1 and a scale height of 6 km. The three images show a high level of consistency considering the differences in spatial resolution. There is no obvious impact of increasing incidences until i = 92° (120 km beyond the terminator), at which stage the S/N ratio becomes marginal as radiance decreases.*

### 3.1. Titan

The observed reflectance (*I/F*) variations of a dark region of Titan with solar zenith angles ranging from 60° to the terminator are shown in Fig. 14. The observed region covers longitudes between 25°E and 55°E at 5°S and appears effectively homogeneous. The observation was taken on September 7 of 2006. Solving for the optical depth, the albedo of the surface and the scale height of aerosols produces a good model fit at all near-IR wavelengths (Fig. 13). The "best-fit" scale height is 80 ± 10 km. This number is compatible with the mean value of 65 km ± 20 km at altitudes higher than 80 km retrieved by Tomasko et al. (2008), particularly considering our use of the simplifying assumption of a constant scale height. The retrieved surface albedo and optical depth spectra are shown in Fig. 14. The retrieved optical depth at 1.08 µm is 2.3 ± 0.5, which is consistent with the value of 2.6 derived from Huygens probe entry dataset of January 2005 (Tomasko et al., 2008). Fitting the retrieved optical depth spectrum with a power law leads to an exponent of 2.2, with the uncertainty allowing a range of 1.8–2.4. This estimate is in good agreement with previous results if we consider the fact that we retrieve a mean optical depth profile for the whole layer. Tomasko et al. (2008) obtained a value of 0.97 below 30 km, 1.41 between 30 km and 80 km and 2.43 above 80 km. Work by Bellucci et al. (2009) inferred values between 1.8 and 2.2 over a large fraction of the altitude range they had analyzed (40–470 km). Once corrected from the contribution of aerosols, the surface albedo spectrum appears relatively flat: it ranges from 0.04 to 0.02, while observed reflectance factors range from 0.18 to 0.07. The uncertainties decrease with wavelength from a relatively high value at 1.0 µm (albedos up to 0.09 also provide a good fit to the observations if associated with a lower optical depth of 1.7) to ±0.005 at 2 µm. In addition, the spectral properties of the surface retrieved with our method at high solar zenith angle are compatible with some of the retrievals and surface analog published by McCord et al. (2006).
.

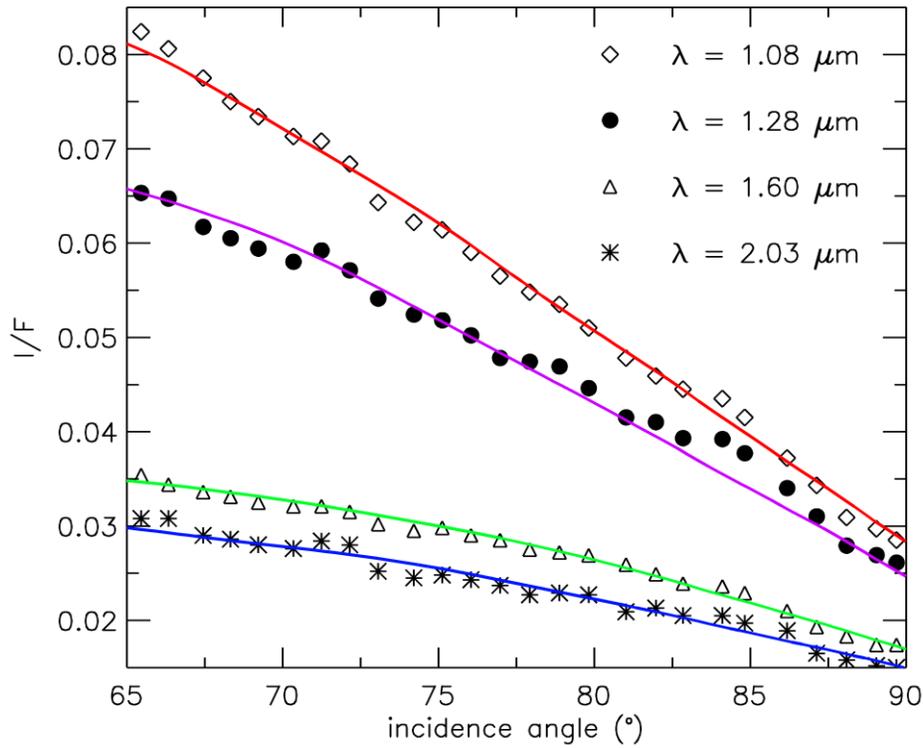

**Figure 13.** *Observed reflectance variations of a dark region of Titan at 5°S, 40°E with solar zenith angle (symbols). Four wavelengths are shown. The best fits of the model to the observation are shown with lines. It is obtained for a scale height of 80 km, and for surface albedos and aerosols optical depths of Fig. 14.*

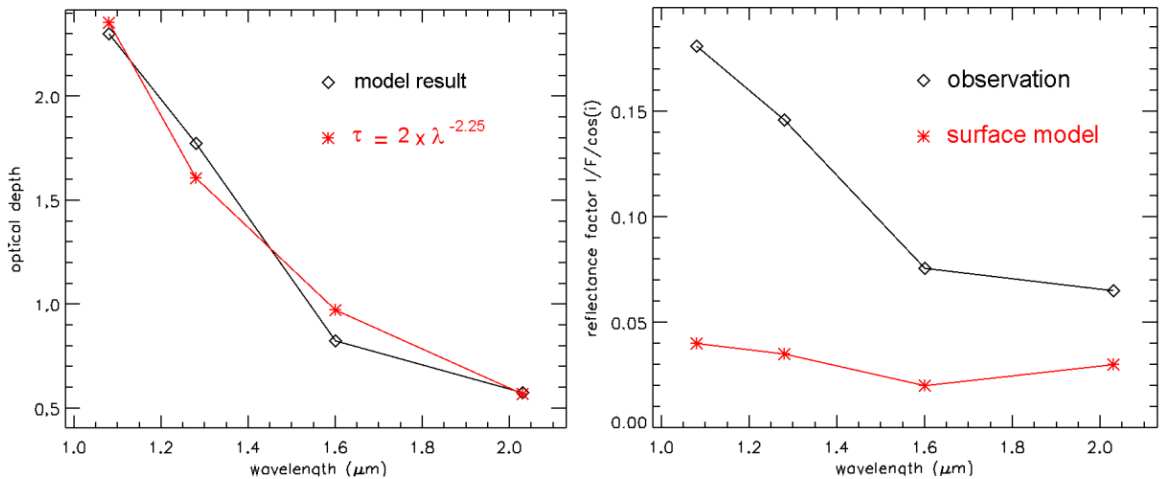

**Figure 14.** *(a) Retrieved optical depth spectrum with the model (black diamond) and best fit of this model with a power law (red stars). (b) Retrieved surface albedo spectrum (red stars) compared to observed reflectance factor (I/F/cos(i)) spectrum (black diamond).*

## 4. Conclusions

We have developed a spherical model of radiative transfer through an "aerosols-rich" atmosphere with the intended purpose of analyzing of remote observations of planetary bodies such as Mars and Titan. Our numerical technique is based on Monte-Carlo methods and therefore does not require any simplifying geometric assumptions in order to solve the radiative transfer. The accuracy is a direct function of the number of photons employed in each simulation, and thus limited only by computational resource issues. Furthermore, we have developed "fast" simulations through the use of pre-calculated look-up tables that contain the required interaction probabilities associated with all possible simple paths in the full 3D spherical atmosphere. As such, the model can be used to rapidly generate simpler look-up tables that are explicit functions of the unknown ("retrievals") parameters associated with a specific set of observations.

This model has been applied to near-IR observations of Mars and Titan acquired by OMEGA/Mars Express and VIMS/Cassini respectively using a sample obtained around the terminator (i.e. at high solar zenith angle or in the night side of the planet) that cannot be analyzed using plane-parallel codes.

Terminator observations are typically quite sensitive to the aerosols properties in the bottom heights above the atmosphere. On Mars, a scale height varying between $6 \pm 2$ km and $11.5 \pm 1.0$ km as a function of the atmospheric temperature is retrieved. The mean particle size of aerosols is inferred to be either constant or to decrease by $0.01$–$0.02$ μm km$^{-1}$. On Titan, an observation obtained on 07/09/2006 is best reproduced using a scale height of $80 \pm 10$ km. At that time the optical depth between 1 μm and 2 μm can be represented by $\tau(\lambda) = 2.3 \times \lambda^{-2.2}$.

In addition to its sensitivity to the vertical properties of aerosols, our model was able to recover surface properties at times and places for which the Sun is low or even below the horizon. This capability is of particular interest for the studies of regions that can only be observed in such non-favorable photometric conditions, such as the highest latitudes of Mars around spring and autumn equinoxes. Considering that the OMEGA and VIMS instruments have acquired numerous observations around the terminator, our model will be able to extend the analysis of the vertical structure of aerosols at different times and locations and to routinely retrieve surface properties up to the terminator and beyond.

**Acknowledgments**

We are grateful to Stéphane Le Mouelic for his help in VIMS data calibration and selection. We would also like to thank Michael J. Wolff and an anonymous Reviewer for useful comments and suggestions.